\documentclass{aa}

\usepackage{graphicx}
\usepackage{supertabular}
\usepackage{epsfig}
\usepackage{natbib}
\usepackage[varg]{txfonts}
\usepackage{hyperref}
\usepackage[normalem]{ulem}
\usepackage{multirow}
\usepackage{longtable}
\usepackage{pdflscape}
\usepackage{color,soul}

\newcommand{\logg}{\ensuremath{\log g}}

\newcommand{\aofe}{\ensuremath{[\mathrm{\alpha/Fe}]}}

\newcommand{\Teff}{\ensuremath{T_{\mathrm{eff}}}}

\newcommand{\beq}{\begin{equation}}
\newcommand{\eeq}{\end{equation}}

\newcommand{\xtmean}[1]{\ensuremath{\left\langle #1\right\rangle}}

\newcommand{\MoH}{\ensuremath{\left[\mathrm{M}/\mathrm{H}\right]}}

\newcommand{\FeoH}{\ensuremath{\left[\mathrm{Fe}/\mathrm{H}\right]}}

\newcommand{\COBOLD}{{\tt CO$^5$BOLD}}
\newcommand{\LHD}{{\tt LHD}}

\newcommand{\ATLAS}{{\tt ATLAS9}}

\newcommand{\MARCS}{{\tt MARCS}}

\newcommand{\LINFOR}{{\tt Linfor3D}}
\newcommand{\MULTI}{{\tt MULTI}}

\newcommand{\twodFH}{{\tt 2dF/HERMES}}
\newcommand{\HERMES}{{\tt HERMES}}

\begin{document}

\title{Abundances of Na, Mg, and K in the atmospheres of red giant branch stars of Galactic globular cluster 47~Tucanae\\}

\author{
   A.\,\v{C}erniauskas\inst{1}
   \and A.\,Ku\v{c}inskas\inst{1}
   \and J.\,Klevas\inst{1}
   \and D.\,Prakapavi\v{c}ius\inst{1}
   \and S.\,Korotin\inst{2,3}
   \and P.\,Bonifacio\inst{4}
   \and H.-G.\,Ludwig\inst{5,4}
   \and E.\,Caffau\inst{4}
   \and M.\,Steffen\inst{6}
   }

\institute{
       Institute of Theoretical Physics and Astronomy, Vilnius University, Saul\.{e}tekio al. 3, Vilnius LT-10222, Lithuania \\
       \email{algimantas.cerniauskas@tfai.vu.lt}
           \and 
       Department of Astronomy and Astronomical Observatory, Odessa National University and Isaac Newton Institute of Chile Odessa
       branch, Shevchenko Park, 65014 Odessa, Ukraine
           \and
           Crimean Astrophysical Observatory, Nauchny 298409, Crimea
           \and
           GEPI, Observatoire de Paris, PSL Research University, CNRS, Place Jules Janssen, 92190 Meudon, France
       \and 
           Zentrum f\"ur Astronomie der Universit\"at Heidelberg, Landessternwarte, K\"onigstuhl 12, 69117 Heidelberg, Germany
           \and
           Leibniz-Institut f\"ur Astrophysik Potsdam, An der Sternwarte 16, D-14482 Potsdam, Germany
           }

\date{Received 21 December 2016 / Accepted 29 March 2017 }

\abstract
   {}
  {We study the abundances of Na, Mg, and K in the atmospheres of 32 red giant branch (RGB) stars in the Galactic globular cluster (GGC) 47~Tuc, with the goal to investigate the possible existence of Na--K and Mg--K correlations/anti-correlations, similar to those that were recently discovered in two other GGCs, NGC~2419 and 2808.}
  {The abundances of K, Na, and Mg were determined using high-resolution \twodFH\ spectra obtained with the Anglo-Australian Telescope (AAT). The one-dimensional (1D) NLTE abundance estimates were obtained using 1D hydrostatic \ATLAS\ model atmospheres and spectral line profiles synthesized with the \MULTI\ package. We also used three-dimensional (3D) hydrodynamical \COBOLD\ and 1D hydrostatic \LHD\ model atmospheres to compute 3D--1D~LTE abundance corrections, $\Delta_{\rm 3D-1D~LTE}$, for the spectral lines of Na, Mg, and K used in our study. These abundance corrections were used to understand the role of convection in the formation of spectral lines, as well as to estimate the differences in the abundances obtained with the 3D hydrodynamical and 1D hydrostatic model atmospheres.}
  {The average element-to-iron abundance ratios and their RMS variations due to star-to-star abundance spreads determined in our sample of RGB stars were $\langle{\rm [Na/Fe]}\rangle^{\rm 1D~NLTE}=0.42\pm0.13$, $\langle{\rm [Mg/Fe]}\rangle^{\rm 1D~NLTE}=0.41\pm0.11$, and $\langle{\rm [K/Fe]}\rangle^{\rm 1D~NLTE}=0.05\pm0.14$. We found no statistically significant relations between the abundances of the three elements studied here. Also, there were no abundance trends with the distance from the cluster center, nor any statistically significant relations between the abundance/abundance ratios and absolute radial velocities of individual stars. All these facts suggest the similarity of K abundance in stars that belong to different generations in 47~Tuc which, in turn, may hint that evolution of K in this particular cluster was unrelated to the nucleosynthesis of Na and/or Mg.}
  {}

        
\authorrunning{\v{C}erniauskas et al.}
\titlerunning{Abundances of K, Na, and Mg in 47~Tuc}

\maketitle

\section{Introduction}

The current paradigm of Galactic globular cluster (GGC) evolution has changed dramatically with the discovery of significant star-to-star variation in the abundances of light elements (such as C, N, O, and Na), as well as correlations/anti-correlations between their abundances \citep[see, e.g.,][for a review]{K94,GSC04}. Photometric studies of numerous GGCs carried out during the past decade have revealed the existence of multiple sub-sequences on their main sequence (MS), subgiant (SGB), and red giant branches (RGB) \citep{PBA07,MPB12}. Further spectroscopic observations have confirmed that stars on these sub-sequences differ in their light element abundances \citep{CBG09a,MVP08,GVL12}, while stars with different light element abundances tend to concentrate in different parts of the cluster \citep{BBC12,CPJ14}. All these observational facts suggest that these stellar systems formed via two or more star formation episodes, thereby producing several generations of stars characterized by different light element abundances and, possibly, also by different kinematical properties \citep[][]{RHA13,KDB14}. This, clearly, contradicts the previously held notion that GGCs are perfect examples of simple stellar populations. 

General consensus today is that during the early evolution of the GGCs, their first-generation (1G) stars somehow enriched second-generation (2G) stars with certain light elements (e.g., Na) and made them depleted in others (e.g., O). Such view is supported, for example, by the existence of various correlations/anti-correlations of light element abundances seen in the GGCs, such as Na--O anti-correlation \citep{CBG06,CBG09a,MVM11}, Mg--Al anti-correlation \citep{CBG09a}, Li--O correlation \citep{PBM05,Shen10}. Also, recent theoretical modeling predicts that stars characterized by different light element abundances (i.e., those that belong to different stellar generations) should have different spatial distributions and dynamical properties that can still be detectable in some GGCs \citep{B11,HGA15} -- which is, again, in line with the observations. Although several competing scenarios have been proposed to explain these observed trends, none of them is able to meet \emph{all} available constraints simultaneously yet \citep[see, e.g.,][]{BLM13,RDC15}.

In this context, the recently discovered Mg--K anti-correlation is especially interesting as it may help shed further light on nucleosynthetic networks that have taken place in stars belonging to the GGCs and, ultimately, may help to discriminate between the different suggested evolutionary scenarios. So far, Mg--K anti-correlation has been observed in two clusters, NGC~2419 \citep[][]{MBI12} and NGC~2808 \citep[][]{MBM15}. Besides the Mg--K anti-correlation, the authors found statistically significant correlations of [K/Fe] with [Na/Fe] and [Al/Fe], and anti--correlation with [O/Fe] abundance ratios. Also, both clusters have extreme He enrichment: NGC~2808 \textit{Y}=0.34 \citep{MMP14},  NGC~2419 \textit{Y}=0.42 \citep{dDM11}. These values are much higher than what has been determined in other globular clusters \citep[see, e.g.,][]{MMD14}. The simultaneous existence of all these trends may suggest that the same self-enrichment mechanisms that produced correlations/anti-correlations between the abundances of other light elements, such as O, Na, Mg, Al, were also responsible for producing the observed spread in K abundance. The question is, of course, whether such correlations/anti-correlations between the abundances K and other light elements can be found in other GGCs. Indeed, it is possible that NGC~2419 and 2808 are not typical examples of genuine GGCs: NGC~2419 may be a remnant of an accreted dwarf galaxy \citep[][]{MV05}, while NGC~2808 shows a very extended O--Na anti-correlation \citep[e.g.,][]{CBG09a} and harbors four Mg-poor (${\rm [Mg/Fe]}<0$) stars \citep[e.g.,][]{CBG09b,C14} -- the properties that are rare amongst other GGCs. In addition, a study of K abundance using small samples of stars in seven ``classical'' GGCs carried out by \citet[][]{CGB13} has revealed no intrinsic scatter in K abundances. Therefore, it is unclear how widespread Mg--K anti-correlation may be amongst the GGCs in general. To answer this question, analysis based on larger stellar samples in other clusters is needed.

In this work we therefore study abundances of Na, Mg, and K in a sample of 32 RGB stars in 47~Tuc with the aim to search for the possible existence of Mg--K anti-correlation and Na--K correlation. This cluster shows interesting connections between chemical abundances and kinematical properties of stars belonging to different generations \citep[e.g.,][]{RHA13,KDB14}. Such connections have not been observed in other GGCs yet. This offers a possibility not only to search for possible connections between the abundance of K and those of Na and/or Mg but to also investigate whether or not stars characterized with different K abundance do also differ in their kinematical properties.

The paper is structured as follows. In Sect.~\ref{sect:method} we present photometric and spectroscopic data used in our study and outline the procedures used to determine 1D~NLTE abundances of Na, Mg, and K, as well as the corresponding 3D--1D abundance corrections. Analysis and discussion of the obtained results is presented in Sect.~\ref{sect:discuss}, while the main findings of this paper and final conclusions are outlined in Sect.~\ref{sect:conclus}.

\section{Methodology}\label{sect:method}

Abundance analysis of Na, Mg, and K was performed in two steps. First, we determined 1D~NLTE abundances using 1D hydrostatic \ATLAS\ model atmospheres and 1D~NLTE abundance analysis methodology. Then, 3D hydrodynamical \COBOLD\ and 1D hydrostatic \LHD\ model atmospheres were used to compute the 3D--1D abundance corrections (see Sect.~\ref{sect:3Dabund}) and to evaluate the influence of convection on the formation of spectral lines used in our study.

\begin{figure}[!t]
        \centering
        \includegraphics[width=9cm]{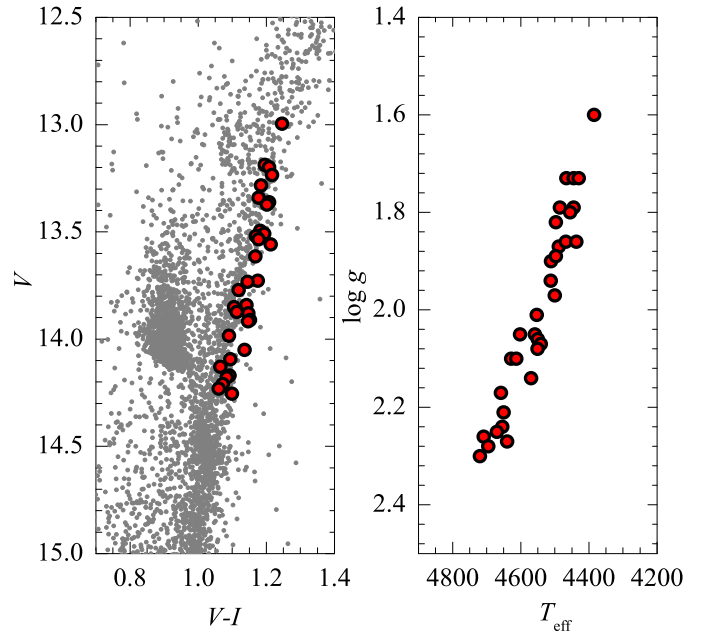}
        \caption{\textbf{Left:} part of $V - (V-I)$ color-magnitude diagram of 47~Tuc from \citet[][]{BS09}. Stars used in this study are marked as filled red circles. \textbf{Right:} a sample of RGB stars used in this work plotted in the $T_{\rm eff} - \log g$ plane.}
        \label{fig:CMD}
\end{figure}

\subsection{Spectroscopic data\label{sect:obs_data}}

In this study we used a sample of 32 RGB stars in 47~Tuc which were observed with the {\tt 2dF} optical fibre positioner and \HERMES\ spectrograph mounted on the Anglo-Australian Telescope (AAT). The raw spectra were obtained during the science verification phase of GALAH survey \citep{DFB15}, and were downloaded from the AAT data {archive}\footnote{\path{http://site.aao.gov.au/arc-bin/wdb/aat_database/observation_log/make}}. We used \twodFH\ spectra obtained in two wavelength regions, 564.9--587.3\,nm ({\tt GREEN}) and 758.5--788.7 ({\tt IR}), using the spectral resolution of $R\sim28000$ and exposure time of 1200\,s. The observations were carried out during the period of Oct 22 -- Dec 20, 2013 (see Table~\ref{tab:obs_log} for details).

\begin{figure*}[!t]
        \centering
        \includegraphics[width=17cm]{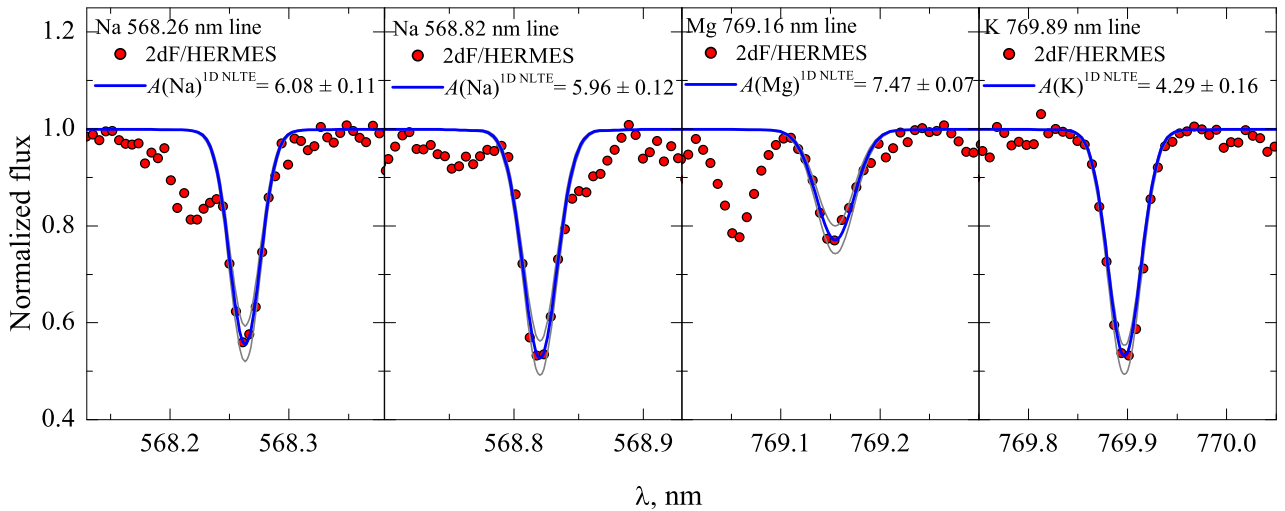}
        \caption{Typical fits of the synthetic Na, Mg, and K line profiles (solid blue lines) to those in the observed {\tt 2dF/HERMES} spectrum (filled red circles) of the target RGB star 47TucS2220 ($\Teff=4490$\,K, $\log g=1.8$). We also indicate the abundances determined from each observed line, $A{\rm (X)}$, together with their errors (see Sec.~\ref{sect:1Dabund}). Thin gray lines show synthetic profiles computed with the abundances altered by $\pm 0.2$\,dex.}
        \label{fig:line_profiles}
\end{figure*}

\begin{table}[]
        \centering
        \caption{Spectroscopic \twodFH\ observations of the target RGB stars in 47~Tuc}\label{tab:obs_log}   
        \begin{tabular}{lccc}
                \hline 
                Date            & \textit{N}  & Exp. time & \\
                                           &     stars    &    s           & \\
                \hline
                2013-10-22 & 1             & 1200        & \\ 
                2013-11-20 & 1             & 1200        & \\
                2013-12-12 & 28            & 1200        & \\
                2013-12-19 & 2             & 1200        & \\
                \hline 
        \end{tabular} 
\end{table}

The raw spectra were reduced using {\tt 2dfdr} data reduction software {(version 6.28)}\footnote{\path{https://www.aao.gov.au/get/document/2dF-AAOmega-obs-manual-71-90.pdf}}. Sky subtraction was carried out using data obtained with 25 sky fibres during each observing night. Continuum normalization was done using the {\tt IRAF} \citep{T86} \textit{continuum} task. Radial velocities of individual stars were measured using the {\tt IRAF} \textit{fxcorr} task, to make sure that all stars are cluster members. The typical signal--to--noise ratio (per pixel) in the vicinity of Na, Mg, and K lines was $S/N\approx50$.

The color-magnitude diagram of 47~Tuc with the target RGB stars marked is shown in Fig.~\ref{fig:CMD}.

\subsection{Atmospheric parameters of the target stars\label{sect:atmos_par}}

Effective temperatures of the sample stars were determined using photometric data from \citet[][]{BS09} and $T_{\rm eff} - (V-I)$ calibration of \citet[][]{RM05}. Prior to the effective temperature determination, photometric data were de-reddened assuming $E(B-V)=0.04$ and the color excess ratio $E(V-I)/E(B-V)=1.33$ \citep[][]{BS09}. Surface gravities were estimated using the classical relation between the surface gravity, mass, effective temperature, and luminosity. Luminosities of individual stars were determined using 12 Gyr Yonsei-Yale isochrones \citep[][]{YDMY01} and absolute $V$ magnitudes obtained assuming the distance modulus $V-M_{\rm{V}}=13.37 $ \citep[][]{H96}. The same set of isochrones was used to estimate masses of the sample stars. Because of a very small mass range occupied by the sample RGB stars ($\sim0.01 {\rm M}_\odot$), identical mass of 0.9 ${\rm M}_\odot$  was adopted for all investigated stars.

\subsection{1D~NLTE abundances of Na, Mg, and K\label{sect:1Dabund}}

Two lines of the same Na doublet transition were used in the determination of Na abundances, 568.26 and 568.82~nm (\HERMES/{\tt GREEN}). In the case of Mg, we used one line located at 769.16~nm (\HERMES/{\tt IR}). Two lines of K resonance doublet located at 766.49\,nm and 769.89~nm were available in the \twodFH\ spectra ({\tt IR} region) of the sample RGB stars. Unfortunately, the weaker line ($\lambda=766.49$\,nm) was strongly blended with telluric ${\rm O}_2$ feature which rendered this K line unusable. For K abundance determination we therefore used the remaining clean line located at 769.89~nm. The two lines of Na and the line of Mg were not affected by telluric lines. 

Atomic data were taken from the VALD-3 database \citep[][]{PKRWJ95,KD11} for the spectral lines of Na and Mg. In case of K, we used oscillator strengths from \citet{M91} and the line broadening constants from the  VALD-3 database. All atomic data are provided in Table~\ref{tab:atom_par} where the line wavelengths, excitation potentials, and oscillator strengths are given in columns 2--4, and the line broadening constants (natural, Stark, and van der Waals) are listed in columns 5--7.

In our 1D~NLTE analysis, we used 1D hydrostatic {\tt ATLAS9} model atmospheres \citep[][]{K93}, which were computed using the Linux port of the {\tt ATLAS9} code \citep[][]{SBC04,S05}. The model atmospheres were calculated using ODFNEW opacity tables \citep[][]{CK03}, with the $\alpha-$element enhancement of $[\alpha/{\rm Fe}]=+0.4$. The mixing length parameter was set to $\alpha_{\rm MLT}=1.25$ and the overshooting was switched off.

\begin{table}[]
        \centering
        \caption{Atomic parameters of Na, Mg, and K lines used in the abundance determinations.}\label{tab:atom_par}    
        \begin{tabular}{lcccccc}
                \hline 
                Element  & $\lambda$, nm & $\chi$, eV & log$\textit{gf}$ & log $\gamma_{rad}$ & log$\frac{\gamma_4}{N_e}$ & log$\frac{\gamma_6}{N_H}$ \\
                \hline
                Na~I       &        568.26      &    2.102     &       $-0.70 $     &              7.84            &                 $-4.22$                &         $-6.85$ \\ 
                Na~I       &        568.82      &    2.104     &       $-0.45 $     &              7.84            &                 $-4.22$                &         $-6.85$ \\ 
                Mg~I      &        769.16      &    5.753     &       $-0.78 $     &              7.57            &                 $-3.25$                &         $-6.83$  \\
                K~I         &        769.89      &    0.000     &       $-0.17 $     &              7.56            &                 $-5.44$                &         $-7.45$  \\ 
                \hline 
        \end{tabular} 
\end{table}

The 1D~NLTE abundances of all three elements were determined using the synthetic spectrum fitting technique. Since the atmospheric parameters of our sample stars were very similar, we used identical microturbulence velocity for all stars, $\xi_{\rm mic}=1.5\,{\rm km~s}^{-1}$. This value was chosen based on the mean microturbulence velocity that was determined in a sample of 139 RGB stars in 47~Tuc, which was constructed by adding together the samples analyzed spectroscopically by \citet[][58 objects]{CBG09a} and \citet[][81 objects]{CPJ14}; we refer to Appendix~\ref{sect_app:err_param} for details. For all sample stars, we also assumed identical rotational velocity of 2\,km~s$^{-1}$ but determined macrotubulence velocities, $\xi_{\rm mac}$ using each individual spectral line. Typical $\xi_{\rm mac}$ values obtained for Na, Mg, and K lines were 3--5\,km~s$^{-1}$. A fixed value of $\FeoH^{\rm 1D~LTE}=-0.76$ \citep[][]{CBG09a} was used throughout this study.

\begin{figure}[!t]
        \centering
        \includegraphics[width=9cm]{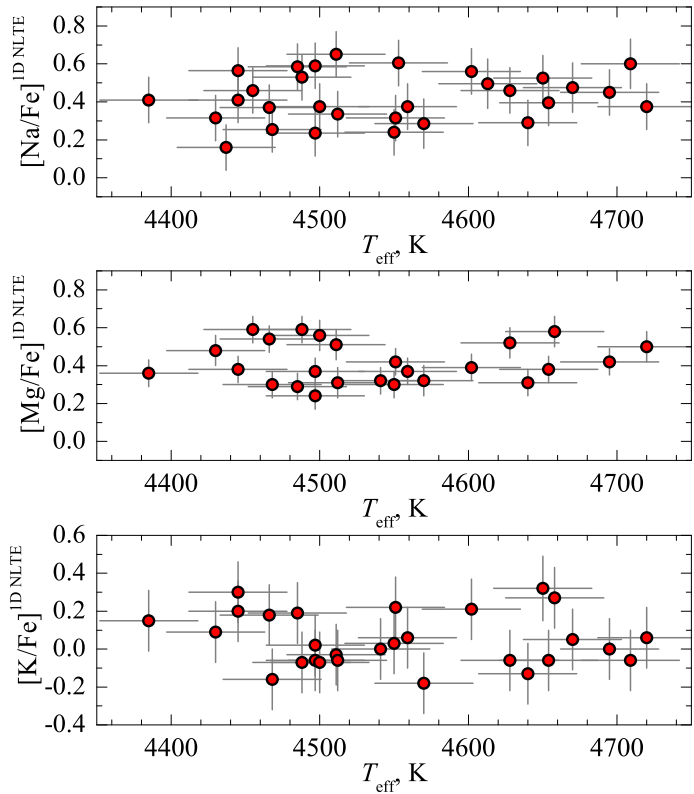}
        \caption{[Na/Fe] (top), [Mg/Fe] (middle), and [K/Fe] (bottom) abundance ratios determined in the sample of 32 RGB stars in 47~Tuc and plotted versus the effective temperature of individual stars.}
        \label{fig:abund-teff}
\end{figure}

The 1D~NLTE abundances of Na, Mg, and K were determined using synthetic line profiles computed with the \MULTI\ code \citep[][]{C86} modified by \citet[][]{KAL99}. The model atom of Na that we used to compute level departure coefficients was presented and thoroughly tested in \citet[][]{DKB14}. It consisted of 20 levels of Na~I and the ground level of Na~II. In total, 46 bound-bound radiative transitions were taken into account when computing level population numbers. 
Collisional cross-sections obtained using quantum mechanical computations \citep[][]{BBDG10} were used for the 9 lowest levels. For other levels, the classical formula of Drawin was utilized in the form suggested by \citet[][]{SH84}, with the correction factor $S_{\rm H} = 1/3$. 
In the case of Mg, we used the model atom from \citet[][]{MSK04}. It consisted of 84 levels of Mg~I, 12 levels of Mg~II, and a ground state of Mg~III. In the computation of departure coefficients, radiative transitions between the first 59 levels of Mg~I and ground level of Mg~II were taken into account. For the 7 lowest levels we used collisional cross-sections obtained using quantum mechanical computations by \citet[][]{BBSG12}, while for other levels we applied the formula of \citet[][]{SH84} with the correction factor $S_{\rm H} = 0.15$. The model atom of K was taken from \citet[][]{ASK10} and consisted of 20 levels of K~I and the ground level of K~II. In addition, another 15 levels of K~I and 7 levels of K~II were used to ensure particle number conservation. The total number of bound-bound radiative transitions taken into account was 62 \citep[see][for further details]{ASK10}. Collisions with neutral H for all levels were taken into account using the formula of \citet[][]{SH84}, with the correction factor $S_{\rm H} = 0.05$ \citep[][]{ASK10}.

\begin{table}[t]
        \begin{center}
                \caption{1D~NLTE--LTE abundance corrections, $\Delta_{\rm 1D~NLTE-LTE}$, for the spectral lines of Na, Mg, and K used in this work (see text for details).}\label{tab:nlte_corr}
                \resizebox{\columnwidth}{!}{
                        \begin{tabular}{lccr@{}lccr@{}lc}
                                \hline\hline
                                Element & Line          && \multicolumn{3}{c}{$\Teff$=4385, $\logg$=1.6}  && \multicolumn{3}{c}{$\Teff$=4720, $\logg$=2.3}  \\ [0.5ex]                              
                                & $\lambda$, nm  && $W$,&~pm    &  $\Delta_{\rm 1D~NLTE-LTE}$ &&  $W$,&~pm  & $\Delta_{\rm 1D~NLTE-LTE}$  \\            
                                \hline 
                                \ion{Na}{i}   & 568.26 && 12&.3 & $-0.25$ && 10&.6 & $-0.19$   \\
                                &                      && 16&.4 & $-0.28$ && 15&.1 & $-0.23$   \\
                                \ion{Na}{i}   & 568.82 && 13&.3 & $-0.26$ && 11&.7 & $-0.20$   \\
                                &                      && 16&.7 & $-0.26$ && 15&.5 & $-0.22$   \\
                                \ion{Mg}{i}   & 769.16 &&  8&.5 & $-0.05$ &&  7&.6 & $-0.01$   \\
                                &                      && 11&.0 & $-0.08$ && 10&.1 & $-0.01$   \\
                                \ion{K}{i}    & 769.89 && 19&.8 & $-0.48$ && 17&.3 & $-0.55$   \\
                                &                      && 24&.0 & $-0.40$ && 21&.6 & $-0.49$   \\
                                \hline
                        \end{tabular}}
                \end{center}
                \vspace{-5mm}
                
        \end{table}

Typical fits of the synthetic 1D~NLTE line profiles to those observed in the \twodFH\ spectrum are shown in Fig.~\ref{fig:line_profiles} while the determined abundances of all three elements are given in Table~\ref{tab_app:all_abund} (Appendix~\ref{sect_app:all_abund}). We verified that for all elements studied there is no dependence of the determined abundances on the effective temperature (Fig.~\ref{fig:abund-teff}). 

Although we directly determined 1D~NLTE abundances of all elements studied here, that is, without measuring their 1D~LTE abundances, we also computed 1D~NLTE--LTE abundance corrections for the spectral lines used in our work. The corrections were calculated using two \ATLAS\ model atmospheres with the effective temperatures and gravities similar to those of stars at the extreme ends of our RGB star sample, for two line equivalent widths for each spectral line used (the $W$ values used represent the minimum and maximum values measured in the RGB star spectra). 
The obtained abundance corrections are provided in Table~\ref{tab:nlte_corr}. Clearly, they are large for the lines of Na and K which demonstrates that NLTE effects play an important role in their formation and that NLTE analysis is necessary in order to obtain reliable abundances using these spectral lines. Indeed, this is in line with the findings of earlier studies (e.g., \citet{TKM09},\citet{LABB11}.

The total errors in the determined abundances of Na, Mg, and K, as well as individual contributions to the total error entering during different steps of the abundance analysis, are given in Table~\ref{tab_app:abund_err} (Appendix~\ref{sect_app:abund_err}; the total error, $\sigma(A)_{\rm tot}$, is a sum of individual contributions in quadratures). The detailed description of how the individual contributions towards the total error were estimated is provided in Appendix~\ref{sect_app:abund_err}.

\subsection{3D+NLTE abundances of Na, Mg, and K\label{sect:3Dabund}}

To study the influence of convection on the formation of \ion{Na}{i}, \ion{Mg}{i}, and \ion{K}{i} lines in the atmospheres of the target RGB stars, we used 3D hydrodynamical \COBOLD\ and 1D hydrostatic \LHD\ model atmospheres \citep[see][respectively]{FSL12,CLS08}. Because the range spanned by the atmospheric parameters of the sample stars was relatively small ($\Delta T\approx300$\,K, $\Delta \log g\approx0.3$), we used 3D hydrodynamical and 1D hydrostatic model atmospheres with the atmospheric parameters similar to those of the median object in our RGB star sample (median star:  $\Teff\approx 4545$\,K and $\log g= 2.02$; model: $\Teff\approx4490$\,K and $\log g=2.0$). Since we did not have models at the metallicity of 47~Tuc, we used those computed at $\MoH=-1.0$. We expect however that the difference between the model and target metallicities should have minor influence on the 3D--1D abundance corrections (see below) because they change little in the interval between $\MoH=-0.7\dots-1.0$ \citep[cf.][]{DKS13}.

The 3D hydrodynamical \COBOLD\ model atmosphere was calculated using \MARCS\ opacities \citep[][]{GEE08} grouped into six opacity bins \citep[see, e.g.,][]{L92,LJS94}. We utilized the solar-scaled abundance table from \citet[][]{GS98}, with CNO abundances set to A(C) = 8.39, A(N) = 7.80,   A(O) = 8.66, and the enhancement in ${\mathrm \alpha}$-element abundances to $\aofe=+0.4$. The \COBOLD\ model atmosphere was computed using a ``box-in-a-star'' setup, on a Cartesian grid of $180\times 180\times 145$ points corresponding to the physical dimensions of $1.65\times1.65\times0.78$\,Gm in two horizontal ($x\times y$) and the vertical ($z$) directions, respectively. Radiative opacities and source function were computed under the assumption of LTE. The entire \COBOLD\ simulation sequence covered the interval of $\approx$\,10 convective turnover times, as measured by the Brunt-Vaias\"{a}l\"{a} timescale \citep[see][for details]{LK12}. 
The 1D hydrostatic \LHD\ model atmosphere was computed using atmospheric parameters, chemical composition, opacities, and equation of state identical to those used in the \COBOLD\ simulation. This choice made it possible to make a strictly differential comparison between the spectral line strengths computed with the two types of model atmospheres.

In order to reduce computational load, spectral line synthesis with the 3D hydrodynamical \COBOLD\ model atmospheres was done using a subset of twenty 3D hydrodynamical model structures computed at different time instants (snapshots) and spaced over the entire span of the model simulation run \citep[see][for details on snapshot selection procedure]{KSL13}. Spectral line synthesis computations were carried out with the \LINFOR\ spectral synthesis package\footnote{\url{http://www.aip.de/Members/msteffen/linfor3d}}. 

The influence of convection on the line strengths was assessed using 3D--1D~LTE abundance corrections, $\Delta_{\rm 3D-1D~LTE}$. The abundance correction measures the difference in the abundance of a given chemical element that would be obtained using 3D hydrodynamical and 1D hydrostatic model atmospheres from the same observed spectral line \citep[e.g.,][]{KSL13,DKS13}. To compute abundance corrections, for each spectral line used in this work we calculated synthetic curves of growth using \COBOLD\ and \LHD\ model atmospheres. Then, for a set of line equivalent widths, $W$, bracketing the values determined for that particular line from the \twodFH\ spectra of different sample stars, we measured the difference in the abundance between the 3D and 1D curves of growth, $\Delta_{\rm 3D-1D~LTE}$. Using this procedure we obtained 3D--1D~LTE abundance corrections at several values of $W$, for each spectral line used in this study (see below).
For computing abundance corrections, instead of using microturbulence velocity utilized in the 1D~NLTE abundance analysis we used $\xi_{\rm mic}$ determined from the 3D hydrodynamical model atmosphere \citep{SCL13}. As argued in \citet[][]{PKD17}, such an approach allows to diminish the influence of shortcomings in the 3D hydrodynamical model atmospheres on the resulting abundance corrections. Microturbulence velocity in the 3D model atmosphere was determined  using Method~1 described in \citet[][]{SCL13}. In this approach, the original velocity field of the 3D hydrodynamical model atmosphere was switched off and the profiles of \ion{Na}{i}, \ion{Mg}{i}, and \ion{K}{i} lines were computed using a set of different depth-independent microturbulence velocities. We then interpolated between the equivalent widths of the obtained line profiles to determine $\xi_{\rm mic}$ at which the equivalent width of the line profile obtained using 3D model and depth independent $\xi_{\rm mic}$ was equal to that computed using full 3D hydrodynamical model atmosphere (i.e., with hydrodynamical velocity field switched on; we note that microturbulence velocities, $\xi_{\rm mic}$, obtained in this way were slightly different for different lines, see Table~\ref{tab:abund_corr}). The obtained value of $\xi_{\rm mic}$ was then used to compute spectral line profiles which were used for constructing 1D~LTE curves-of-growth. These curves-of-growth, together with those computed using full 3D hydrodynamical model atmosphere, were used to calculate 3D--1D~LTE abundance corrections, $\Delta_{\rm 3D-1D~LTE}$.

\begin{table}
        \centering
        \caption{The 3D--1D abundance corrections, $\Delta_\mathrm{3D-1D~LTE}$, for the different strengths of spectral lines of \ion{Na}{i}, \ion{Mg}{i}, and \ion{K}{i} used in this work (see text or details).}\label{tab:abund_corr}
        \centering
        \begin{tabular}{lcccccc}
                \hline\hline 
                Element      & $\lambda_{\rm central}$ & $\xi_\mathrm{micro}$ & \multicolumn{3}{c}{$\Delta_\mathrm{3D-1D~LTE}$, dex} \\
                                  &                nm                   &        km s$^{-1}$         & weak  & strong                                  \\
                \hline
                \ion{Na}{i}  & 568.3 nm                        &     $1.11\pm0.02$       & +0.05 & +0.06 \\
                \ion{Na}{i}  & 568.8 nm                        &     $1.11\pm0.02$       & +0.06 & +0.07 \\
                \ion{Mg}{i} & 769.2 nm                         &     $1.03\pm0.01$       & +0.06 & +0.07 \\
                \ion{K}{i}    & 769.9 nm                           &     $1.06\pm0.002$      & +0.05 & +0.05 \\
                \hline 
        \end{tabular}
\end{table}

\begin{figure*}[!t]
        \centering
        \mbox{\includegraphics[width=17cm]{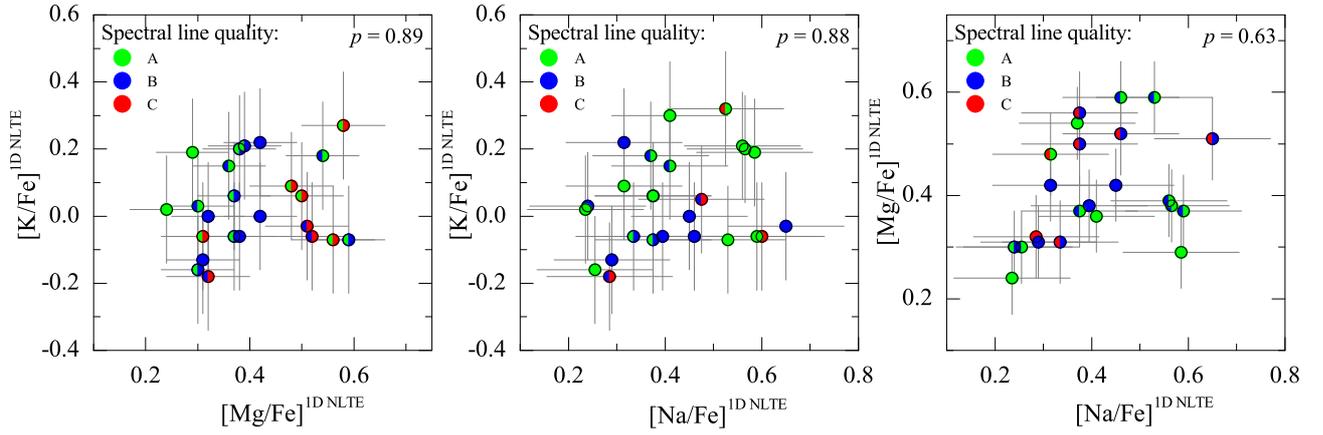}}
        \caption{ Abundances of Na, Mg, and K determined in the sample of RGB stars and shown in various abundance-abundance planes. Colors on the left and right sides of the symbols correspond to the quality (class) of lines used to determine abundances plotted on the $y$ and $x$ axes, respectively (see Appendix~\ref{sect_app:quality} for details). The two-tailed probability $p$-values are marked in the corresponding panels.}
        \label{fig:abund-ratios}
\end{figure*}

\begin{figure*}[!t]
        \centering
        \mbox{\includegraphics[width=7.5cm]{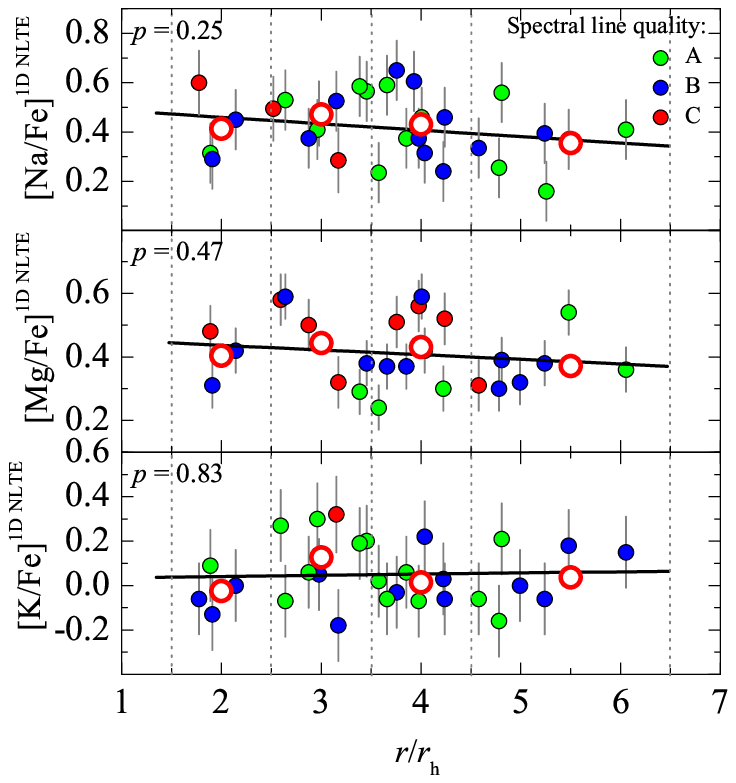}}
        \mbox{\includegraphics[width=7.5cm]{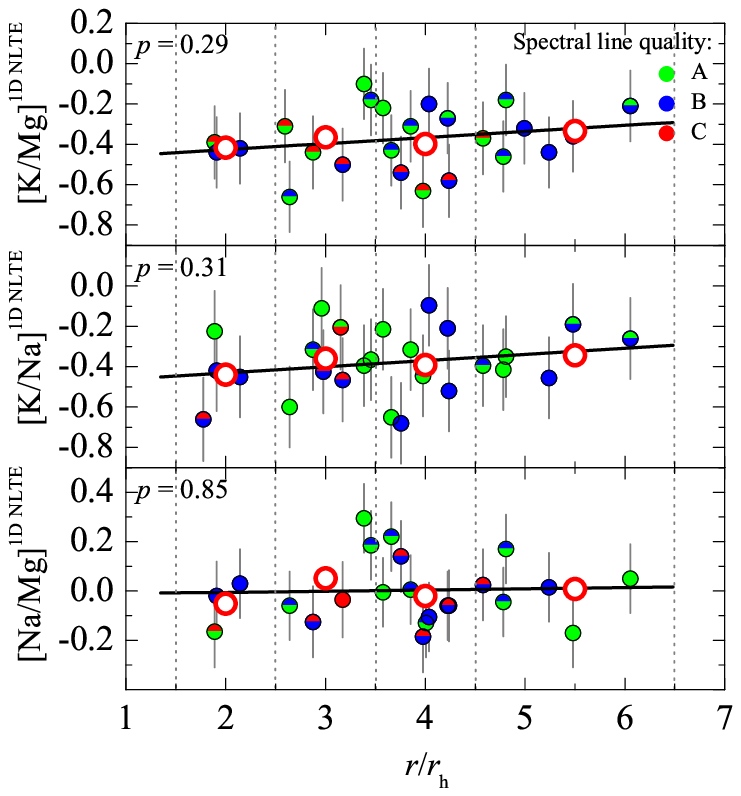}}
        \caption{ Abundances of Na, Mg, and K (left) and abundance-abundance ratios (right) plotted versus the distance from the cluster center, $r/r_{\rm h}$ (small filled circles; $r_{\rm h}$ is a half-mass radius of 47~Tuc, $r_{\rm h}=174^{\prime\prime}$, taken from \citealt{TDK93}). Symbol color denotes quality (class) of the spectral lines from which the abundance was determined (see Fig.~\ref{fig:abund-ratios}). Large open circles are average abundances/abundance ratios computed in non-overlapping $\Delta r/r_{\rm h}=1$ wide distance bins (marked by the vertical dashed lines; note that the outermost bin is wider; error bars measure the RMS scatter of abundances/abundance ratios in a given bin). Black solid lines are linear fits to the data of individual stars, with the $p$-values (see text) marked in the corresponding panels.}
        \label{fig:abund-radial}
\end{figure*}

\begin{figure*}[!t]
        \centering
        \mbox{\includegraphics[width=7cm]{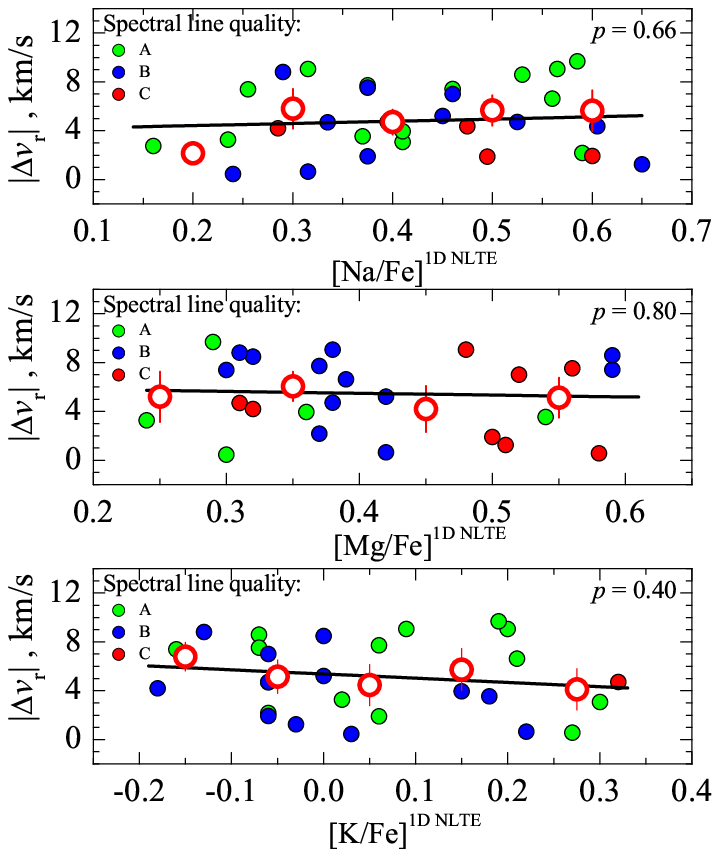}}
        \mbox{\includegraphics[width=7cm]{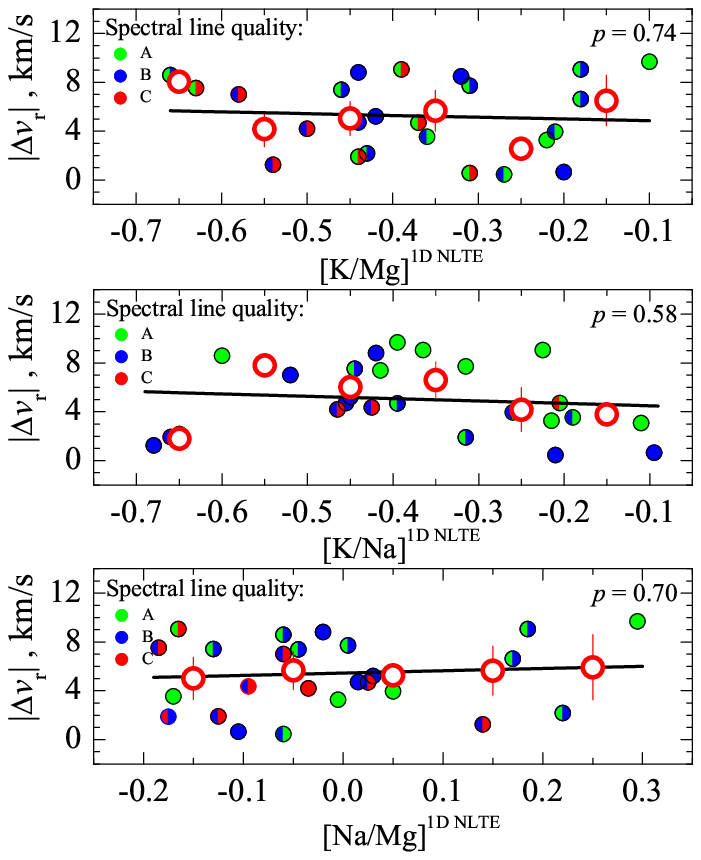}}
        \caption{Absolute radial velocities of RGB stars in 47~Tuc, $\left| \Delta v_{\rm r}\right| $, plotted versus the abundance (left) and  abundance-abundance ratios (right) of Na, Mg, and K. Measurements of individual stars are shown as small filled circles while the absolute radial velocities averaged in 0.1\,dex-wide non-overlapping abundance bins are plotted as large open circles (error bars in the latter case measure RMS scatter of $\left| \Delta v_{\rm r} \right |$ in a given abundance bin). Symbol color denotes quality (class) of the spectral lines from which the abundance was determined (see Fig.~\ref{fig:abund-ratios}). Solid lines are linear fits to the data of individual stars, with the $p$-values (see text) marked in the corresponding panels.}
        \label{fig:disper-radial}
\end{figure*}

The obtained 3D--1D~LTE abundance corrections are provided in Table~\ref{tab:abund_corr} for all lines of Na, Mg, and K used in this study. Since the abundance corrections showed little variation with the line strength, in the case of each element they are provided for two values of $W$ (corresponding to ``weak'' and ``strong'' spectral lines) that bracket the range of equivalent widths measured in the observed spectra of the sample RGB stars. The following line strengths were used: for the weakest lines 8.5, 5, and 20~pm, and for the strongest lines 14, 9, and 22~pm, in case of Na, Mg, and K, respectively. We also provide microturbulence velocities determined for each spectral line using 3D hydrodynamical \COBOLD\ model atmosphere, as described above, as well as the range in $\xi_{\rm mic}$ computed at the extreme values of $W$ (Table~\ref{tab:abund_corr}, column~3). For all lines, the abundance corrections were small and did not exceed $\approx0.07$\,dex. 

Based on the small size of the abundance corrections, we therefore conclude that convection plays a minor role in the formation of these particular spectral lines in the atmospheres of RGB stars in 47~Tuc. This is in line with our earlier findings which have shown that at the metallicity of 47~Tuc the $\Delta_{\rm 3D-1D~LTE}$ corrections expected for lines of \ion{Na}{i}, \ion{Mg}{i}, and \ion{K}{i} should be small, typically, $<0.10$\,dex (\citealt[][]{DKS13}; see also \citealt[][]{CAT07}). Obviously, NLTE effects may play their part in, for example, changing the concentration of neutral atoms in the red giant atmospheres due to overionization. This may have an impact on the line strengths and, thus, the abundance corrections. At the moment, however, we are not able to evaluate the size of these effects for Na, Mg, and K in 3D~NLTE. Nevertheless, in the case of \ion{Li}{i}, which has an atomic structure similar to that of \ion{Na}{i} and \ion{K}{i}, these effects are typically small, with the resulting influence on the abundance corrections of $<0.1$\,dex \citep[][]{KKS16}. 

One may therefore conclude that $\Delta_{\rm 3D-1D~LTE}$ abundance corrections were found to be small and it is likely that they will not be dramatically different if obtained in 3D~NLTE. Nevertheless, we refrain from adding the obtained 3D--1D~LTE corrections to our 1D~NLTE abundances. As it was shown in \citet[][]{KKS16} that such procedure is generally incorrect since it may predict abundances that are different from those that would be obtained using the full 3D~NLTE approach, especially at sub-solar metallicities. In any case, given their size, the 3D--1D abundance corrections are not very important in the context of the present study as applying them would only result in a small uniform shift of the abundance zero-points.

\section{Results and discussion}\label{sect:discuss}

Abundances of potassium in 47~Tuc were investigated by \citet{CGB13}. The authors used three turn--off (TO) and nine SGB stars and obtained the average 1D~NLTE element-to-iron abundance ratios $\langle{\rm [K/Fe]}\rangle_{\rm TO}=0.19\pm0.07$ and $\langle{\rm [K/Fe]}\rangle_{\rm SGB}=0.12\pm0.12$, respectively (the error here is RMS deviation due to star--to--star abundance  variation; for simplicity, we henceforth refer to the element-to-iron abundance ratio as ``abundance'' because, for computing these ratios, we used a fixed iron abundance, $\FeoH^{\rm 1D~LTE}=-0.76$ (see see Sect.~\ref{sect:1Dabund}) and thus the star-to-star variation in, for example,  the [K/Fe] abundance ratio is caused solely by the variation in K abundance). These values are comparable with the average K abundances obtained in this work using RGB stars, $\langle{\rm [K/Fe]}\rangle^{\rm 1D~NLTE}=0.05\pm0.14$, although the star-to-star variation is slightly larger in our case. In case of Na and Mg, the average values obtained by us are $\langle{\rm [Na/Fe]}\rangle^{\rm 1D~NLTE}=0.42\pm0.13$ and $\langle{\rm [Mg/Fe]}\rangle^{\rm 1D~NLTE}=0.41\pm0.11$, respectively. Again, these abundances are comparable with those obtained in 1D~NLTE by \citet[][]{CBG09b}, $\langle{\rm [Na/Fe]}\rangle=0.53\pm0.15$ and $\langle{\rm [Mg/Fe]}\rangle=0.52\pm0.03$ though scatter in the case of Mg is significantly larger in our case (also see below).

Recently, a study of potassium abundance in 47~Tuc was presented in \citet{MMB17}. Authors derived 1D NLTE abundance of potassium in 144 RGB stars and obtained the average element-to-iron abundance ratio $\langle{\rm [K/Fe]}\rangle_{\rm RGB}=-0.12\pm0.01$. This value is 0.17 dex lower than the one obtained in our study. Nevertheless, the difference between the two abundance estimates is not very large and can be explained by differences in the analysis techniques, for example, different approaches with treating microturbulence velocities in the two studies (a more exhaustive comparison with these results will be presented in our forthcoming paper on K abundance in 47~Tuc). It is important to stress, however, that, just as in our study, \citet{MMB17} find no intrinsic spread of potassium abundance in 47~Tuc.

The 1D~NLTE abundances of Na, Mg, and K obtained in our study are plotted in Fig.~\ref{fig:abund-ratios} (the abundances shown in this plot were computed using a fixed value of $\FeoH^{\rm 1D~LTE}=-0.76$, see Sect.~\ref{sect:1Dabund}). One may notice a significant star-to-star variation in the case of all three elements. In order to investigate whether or not this variation may be caused by intrinsic spread in the elemental abundances, we used the maximum-likelihood (ML) technique identical to the one applied by \citet[][]{MBI12,MBM15} to study K abundance spreads in NGC~2419 and NGC~2808 (see Appendix~\ref{sect_app:ML_test} for details). The results of the ML test suggest that there may exist small intrinsic abundance spreads in the case of Na and Mg; 0.04\,dex and 0.08\,dex, respectively (Table~\ref{tab_app:ML_values}). In fact, in the case of Mg the abundance determination errors may be larger than those provided in Table~\ref{tab_app:abund_err} (see Sect.~\ref{sect_app:abund_err}). This may be also indirectly supported by the larger total star-to-star RMS abundance variation seen in our sample, for example, in comparison to that determined in other studies \citep[e.g.,][]{CBG09b}. Therefore, the existence of intrinsic abundance spread in the case of Mg still needs to be confirmed. However, we detected no intrinsic spread in the abundance of K.

We find no statistically significant correlations in different abundance-abundance planes shown in Fig.~\ref{fig:abund-ratios}. For this, we verified whether or not the null hypothesis (i.e., that the slope in the given plane is zero) can be rejected based on the two-tailed probability $p$ that was determined from our data using Student's $t$-test (with smaller $p$ meaning higher evidence against the null hypothesis). In all planes the obtained $p$ values were $p\geq63$\%, thus rendering the rejection of the null hypothesis unwarranted. Similarly, no statistically significant relations were found between either the  abundance or different abundance-abundance ratios of light elements measured in the atmosphere of a given star and the star's distance from the cluster center (the data is shown in Fig.~\ref{fig:abund-radial}): in all planes we obtained $p\geq25$\%. In the case of the [Na/Fe] ratio, non-detection agrees with our earlier result obtained using 102 TO stars in 47~Tuc where no correlation between Na  abundance and distance from the cluster center was found \citep[cf.][]{KDB14}.

Following \citet[][]{KDB14}, we also computed absolute radial velocities of the target RGB stars, $\left| \Delta v_{\rm r}\right| \equiv \left| v_{\rm rad} - {\langle v_{\rm rad} \rangle}^{\rm clust} \right|$, where $v_{\rm rad}$ is radial velocity of the individual star and ${\langle v_{\rm rad} \rangle}^{\rm clust} = -18.6$\,km/s is the mean radial velocity of the sample. The obtained absolute radial velocities of individual stars are plotted versus abundance and different abundance-abundance ratios in Fig.~\ref{fig:disper-radial} (we also show the average absolute velocities which were computed in 0.1\,dex-wide abundance bins). For all elements, the obtained \textit{p} values were $p\geq40$\% thereby rendering the rejection of the null hypothesis unwarranted.

Taken together, these results suggest that stars belonging to different stellar generations in 47~Tuc (as judged, e.g., by their Na abundance) do not show statistically significant differences in the abundances of Mg and/or K. The obtained results also suggest that different stellar generations do not differ in their mean absolute radial velocities. The latter finding contradicts our results obtained using 102 TO stars in 47~Tuc which suggested that kinematical properties of stars in primordial (1G) and extreme generations (3G) may be different \citep[][]{KDB14}. It is important to note though that the sample of RGB stars used in this study is kinematically cooler than that of TO stars used in our previous study. For example, the fraction of stars with $\left| \Delta v_{\rm r}\right| \leq 8.0$\,km/s is larger in our sample than that in the sample of TO stars, $\approx83\%$ versus $\approx70\%$, respectively. Also, in our sample there are no stars with $\left| \Delta v_{\rm r}\right| > 10.0$\,km/s while the fraction of such objects in the TO sample is $\approx18\%$. Therefore, it is possible that this disagreement between the findings of the two studies may be due to the fact that in our sample we miss part of the primordial generation (1G) stars characterized by the largest absolute radial velocities.

\section{Conclusions}\label{sect:conclus}

We studied abundances of Na, Mg, and K in the atmospheres of 32 RGB stars that are members of Galactic globular cluster 47~Tuc. The abundances were determined using archival \twodFH\ spectra that were obtained with the Anglo-Australian Telescope in two wavelength regions, 564.9--587.3\,nm ({\tt GREEN}) and 758.5--788.7 ({\tt IR}). The spectroscopic data were analyzed using 1D \ATLAS\ model atmospheres and 1D~NLTE abundance analysis methodology. In the case of all three elements, 1D~NLTE spectral line synthesis was performed with the \MULTI\ package, using up-to-date model atoms of Na, Mg, and K. We also used 3D hydrodynamical \COBOLD\ and 1D hydrostatic \LHD\ model atmospheres to compute 3D--1D abundance corrections for the spectral lines of all three elements utilized in this study (3D--1D abundance correction is a difference in the abundance of a given element that would be obtained from the same spectral line with the 3D hydrodynamical and 1D hydrostatic model atmospheres). The obtained abundance corrections were small, in all cases $\leq0.07$\,dex, indicating that the influence of convection on the formation of these spectral lines in the atmospheres of RGB stars in 47~Tuc is minor. 

The obtained sample-averaged abundances were $\langle{\rm [Na/Fe]}\rangle^{\rm 1D~NLTE}=0.42\pm0.13$, $\langle{\rm [Mg/Fe]}\rangle^{\rm 1D~NLTE}=0.41\pm0.11$, and $\langle{\rm [K/Fe]}\rangle^{\rm 1D~NLTE}=0.05\pm0.14$ (numbers after the $\pm$ sign are RMS abundance variations due to star-to-star scatter). These numbers agree well with the average abundances obtained earlier by other authors \citep[e.g.,][]{CBG09b, CGB13}. However, we detected somewhat larger star-to-star variation in the case of Mg which may be due to larger systematical uncertainties in its abundance determination. 

We found no statistically significant relations between the abundances of Na, Mg, and K (we note that we actually compared element-to-iron abundance ratios though in all cases we used identical fixed iron abundance). We also detected no significant correlations/anti-correlations between the  abundance/abundance-abundance ratios and distance from the cluster center. Finally, there were no relations between the absolute radial velocities of individual stars and abundances of the light elements in their atmospheres. These results may suggest that production of Na and K, and that of Mg and K has evolved via different and, possibly, unrelated pathways in 47~Tuc. Nevertheless, we stress that RGB stars studied in this work are slightly cooler kinematically than TO stars used in our earlier study where we detected differences in the absolute kinematical velocities of stars in different generations \citep[][]{KDB14}. As a consequence, some stars with the largest absolute radial velocities, which according to \citet{KDB14} should be predominantly those that belong to primordial generation (1G), may be missing in our sample of RGB stars. Therefore, while our present results suggest that the origin of K in 47~Tuc has not been directly related to the nucleosynthesis of Na and Mg, it still needs to be verified whether this conclusion may also hold true in the case of stars characterized by the largest absolute radial velocities.

\begin{acknowledgements}
{Based on data acquired through the Australian Astronomical Observatory, under program 2013B/13. This work was supported by a grant from the Research Council of Lithuania (MIP-089/2015). HGL acknowledges financial support by Sonderforschungsbereich SFB 881 ``The Milky Way System'' (subproject A4) of the German Research Foundation (DFG). AC is grateful to V.~Dobrovolskas for his help with the technical issues of the manuscript preparation. We also thank the anonymous referee for her/his constructive comments and suggestions that have helped to improve the paper.}
\end{acknowledgements}


\bibliographystyle{aa}

\begin{appendix}
        
\section{Abundances of Na, Mg, and K determined in the sample of RGB stars in 47~Tuc}\label{sect_app:all_abund}
        
In this Section we provide a list of target RGB stars (32 objects), their atmospheric parameters (determined in Sect.~\ref{sect:atmos_par}), and abundances of Na, Mg, and K (Sect.~\ref{sect:1Dabund} and \ref{sect:3Dabund}). This information is summarized in Table~\ref{tab_app:all_abund} the contents of which are as follows:
        
\begin{itemize} 

        \item column~1: object name;
        \item column~2: right ascension;
        \item column~3: declination;
        \item column~4: $V$ magnitude;
        \item column~5: $I$ magnitude;
        \item column~6: effective temperature;
        \item column~7: surface gravity;
        \item column~8: radial velocity;
        \item column~9: 1D NLTE sodium abundance and its error;
        \item column~10: 1D NLTE magnesium abundance and its error;
        \item column~11: 1D NLTE potassium abundance and its error.
        
\end{itemize}

\onecolumn

\begin{center}
\begin{longtable}{lccccccr@{}lccr@{}l}
\caption{Target RGB stars in 47~Tuc, their atmospheric parameters, and determined abundances of Na, Mg and K.}
\label{tab_app:all_abund}\\                                   
\hline\hline
\smallskip  
Star         & RA       & Dec.    & $V$ & $I$ & \Teff   & $\log g$ &  \multicolumn{2}{c}{$v_{\rm rad}$} & ${\rm[Na/Fe]}$ &  ${\rm[Mg/Fe]}$   & \multicolumn{2}{c}{${\rm[K/Fe]}$}  \\
name         & (2000)   & (2000)  & mag & mag &  K      & [cgs]    &   km  &   /s                       & { }1D NLTE     &   1D NLTE         & 1D &{ }NLTE                      \\
\hline      
\vspace{-0.2 cm}
\endfirsthead
\hline      
\vspace{-0.2 cm}
\endhead    
\hline      
\endfoot    
N104-S1667   & 6.40433 & --72.20061 &  13.19 & 11.99 & 4450 & 1.73 &  --9&.5 & 0.56 $\pm$0.12 &  0.38 $\pm$0.07 &   0.&20 $\pm$0.16 \\
N104-S2214   & 6.31979 & --72.07552 &  13.23 & 12.01 & 4430 & 1.73 & --27&.6 & 0.31 $\pm$0.12 &  0.48 $\pm$0.08 &   0.&09 $\pm$0.16 \\
N104-S1213   & 5.79479 & --71.93364 &  13.28 & 12.09 & 4490 & 1.79 &  --8&.9 & 0.58 $\pm$0.12 &  0.29 $\pm$0.07 &   0.&19 $\pm$0.16 \\
N104-S1430   & 5.98271 & --71.90491 &  13.34 & 12.16 & 4500 & 1.82 & --16&.4 & 0.59 $\pm$0.12 &  0.37 $\pm$0.07 & --0.&06 $\pm$0.16 \\
N104-S1625   & 6.35437 & --71.98103 &  13.36 & 12.15 & 4450 & 1.79 & --21&.6 & 0.41 $\pm$0.12 &  ...            &   0.&30 $\pm$0.16 \\
N104-S2499   & 6.64554 & --72.11236 &  13.37 & 12.17 & 4460 & 1.80 & --11&.2 & 0.46 $\pm$0.12 &  0.59 $\pm$0.07 &   &{ }{ }{ }{ }...\\
N104-S2220   & 6.36883 & --72.01095 &  13.49 & 12.31 & 4490 & 1.87 & --27&.2 & 0.53 $\pm$0.12 &  0.59 $\pm$0.07 & --0.&07 $\pm$0.16 \\
N104-S1800   & 6.55300 & --72.00191 &  13.52 & 12.35 & 4510 & 1.90 & --17&.3 & 0.65 $\pm$0.12 &  0.51 $\pm$0.08 & --0.&03 $\pm$0.16 \\
N104-S1481   & 6.09350 & --72.25283 &  13.53 & 12.35 & 4500 & 1.89 & --21&.8 & 0.23 $\pm$0.12 &  0.24 $\pm$0.07 &   0.&02 $\pm$0.16 \\
N104-S1849   & 6.60921 & --72.01517 &  13.72 & 12.55 & 4500 & 1.97 & --26&.1 & 0.38 $\pm$0.12 &  0.56 $\pm$0.08 & --0.&07 $\pm$0.16 \\
N104-S1779   & 6.53142 & --71.97400 &  13.73 & 12.58 & 4550 & 2.01 & --22&.9 & 0.61 $\pm$0.12 &  ...            &   &{ }{ }{ }{ }...\\
N104-S1490   & 6.13163 & --72.26456 &  13.84 & 12.69 & 4560 & 2.05 & --10&.8 & 0.38 $\pm$0.12 &  0.37 $\pm$0.07 &   0.&06 $\pm$0.16 \\
N104-S2328   & 5.39333 & --72.14819 &  13.85 & 12.74 & 4630 & 2.10 & --11&.5 & 0.46 $\pm$0.12 &  0.52 $\pm$0.08 & --0.&06 $\pm$0.16 \\
N104-S1636   & 6.36917 & --72.02222 &  13.87 & 12.76 & 4610 & 2.10 & --16&.7 & 0.49 $\pm$0.13 &  ...            &   &{ }{ }{ }{ }...\\
N104-S1751   & 6.51550 & --72.20506 &  13.91 & 12.77 & 4550 & 2.08 & --19&.2 & 0.31 $\pm$0.12 &  0.42 $\pm$0.07 &   0.&22 $\pm$0.16 \\
N104-S1657   & 6.39483 & --72.13347 &  13.98 & 12.89 & 4660 & 2.17 & --18&.0 & ...            &  0.58 $\pm$0.08 &   0.&27 $\pm$0.16 \\
N104-S1750   & 6.51125 & --72.05508 &  14.09 & 13.00 & 4650 & 2.21 & --23&.3 & 0.53 $\pm$0.12 &  ...            &   0.&32 $\pm$0.17 \\
N104-S1543   & 6.23467 & --72.13755 &  14.12 & 13.06 & 4710 & 2.26 & --16&.6 & 0.60 $\pm$0.13 &  ...            & --0.&06 $\pm$0.16 \\
N104-S1732   & 6.49112 & --72.08711 &  14.17 & 13.09 & 4670 & 2.25 & --22&.9 & 0.47 $\pm$0.13 &  ...            &   0.&05 $\pm$0.16 \\
N104-S1563   & 6.27129 & --72.15166 &  14.21 & 13.13 & 4700 & 2.28 & --13&.4 & 0.45 $\pm$0.12 &  0.42 $\pm$0.07 &   0.&00 $\pm$0.16 \\
N104-S167    & 5.09500 & --72.02045 &  12.99 & 11.75 & 4390 & 1.60 & --22&.5 & 0.41 $\pm$0.12 &  0.36 $\pm$0.07 &   0.&15 $\pm$0.16 \\
N104-S292    & 5.17163 & --72.04539 &  13.18 & 11.99 & 4470 & 1.73 & --15&.0 & 0.37 $\pm$0.12 &  0.54 $\pm$0.07 &   0.&18 $\pm$0.16 \\
N104-S1084   & 5.60667 & --71.88953 &  13.50 & 12.31 & 4470 & 1.86 & --11&.2 & 0.25 $\pm$0.12 &  0.30 $\pm$0.07 & --0.&16 $\pm$0.16 \\
N104-S2474   & 6.47350 & --71.86870 &  13.55 & 12.34 & 4440 & 1.86 & --15&.8 & 0.16 $\pm$0.12 &  ...            &   &{ }{ }{ }{ }...\\
N104-S1844   & 6.60458 & --71.95181 &  13.61 & 12.44 & 4510 & 1.94 & --23&.2 & 0.33 $\pm$0.12 &  0.31 $\pm$0.08 & --0.&06 $\pm$0.16 \\
N104-S2333   & 5.40721 & --72.21652 &  13.77 & 12.65 & 4600 & 2.05 & --25&.2 & 0.56 $\pm$0.12 &  0.39 $\pm$0.07 &   0.&21 $\pm$0.16 \\
N104-S1070   & 5.58192 & --72.23414 &  13.87 & 12.73 & 4550 & 2.06 & --18&.1 & 0.24 $\pm$0.12 &  0.30 $\pm$0.07 &   0.&03 $\pm$0.16 \\
N104-S2494   & 6.62525 & --72.23708 &  13.91 & 12.75 & 4540 & 2.07 & --10&.1 & ...            &  0.32 $\pm$0.07 &   0.&00 $\pm$0.16 \\
N104-S1692   & 6.43354 & --71.99447 &  14.05 & 12.91 & 4570 & 2.14 & --22&.8 & 0.29 $\pm$0.13 &  0.32 $\pm$0.08 & --0.&18 $\pm$0.16 \\
N104-S2439   & 6.24533 & --71.83745 &  14.17 & 13.07 & 4650 & 2.24 & --23&.3 & 0.39 $\pm$0.12 &  0.38 $\pm$0.07 & --0.&06 $\pm$0.16 \\
N104-S1524   & 6.19971 & --72.20933 &  14.23 & 13.17 & 4720 & 2.30 & --16&.7 & 0.38 $\pm$0.12 &  0.50 $\pm$0.08 &   0.&06 $\pm$0.16 \\
N104-S2176   & 6.21292 & --72.00975 &  14.25 & 13.15 & 4640 & 2.27 & --27&.4 & 0.29 $\pm$0.12 &  0.31 $\pm$0.07 & --0.&13 $\pm$0.16 \\
\hline                                                                                
mean         &         &            &        &       &      &      &     &   & 0.42           & 0.41            &     &{ }{ }0.05   \\
$\sigma$     &         &            &        &       &      &      &     &   & 0.13           & 0.11            &     &{ }{ }0.14    \\                                                   

\end{longtable}                                          
\end{center}

\twocolumn

\section{Errors in the abundances of Na, Mg, and K}\label{sect_app:abund_err}

In this study we applied a maximum-likelihood technique in order to estimate the extent of possible intrinsic star-to-star variation in the abundances of Na, Mg, and K. For this, we needed realistic estimates of the abundance errors. The procedures used in their derivation are briefly summarized below. Note that in all cases these errors represent the lower limits since they do not account for various possible systematic errors.

\begin{table*}[t]
        \begin{center}
                \caption{Errors in the determined abundances of Na, Mg, and K.}
            \label{tab_app:abund_err}
                \begin{tabular}{lccccccccc}
                        \hline\hline
                        Element & Line & Line     &  $\sigma(\Teff)$ & $\sigma(\log g)$  &      $\sigma(\xi_{\rm t})$    &  $\sigma(\rm cont)$ &  $\sigma(\rm fit)$  & $\sigma(A)_{\rm tot}$  \\ [0.5ex]
                        & wavelength, nm          & quality  &             dex              &             dex           &        dex         &   dex    &    dex    &   dex             \\
                        \hline
                        \ion{Na}{i}   & 568.26 & A & $\pm0.07$ & $\mp0.01$  & $\mp0.08$ & $0.03$  & $0.03$ &  0.11 \\
                        &                      & B & $\pm0.07$ & $\mp0.01$  & $\mp0.08$ & $0.03$  & $0.04$ &  0.12 \\
                        &                      & C & $\pm0.07$ & $\mp0.01$  & $\mp0.08$ & $0.03$  & $0.04$ &  0.12 \\
                        \ion{Na}{i}   & 568.82 & A & $\pm0.07$ & $\mp0.01$  & $\mp0.08$ & $0.04$  & $0.02$ &  0.12 \\
                        &                      & B & $\pm0.07$ & $\mp0.01$  & $\mp0.08$ & $0.04$  & $0.03$ &  0.12 \\
                        &                      & C & $\pm0.07$ & $\mp0.01$  & $\mp0.08$ & $0.04$  & $0.06$ &  0.13 \\
                        \ion{Mg}{i}   & 769.16 & A & $\pm0.03$ & $\mp0.01$  & $\mp0.05$ & $0.03$  & $0.02$ &  0.07 \\
                        &                      & B & $\pm0.03$ & $\mp0.01$  & $\mp0.05$ & $0.03$  & $0.03$ &  0.07 \\
                        &                      & C & $\pm0.03$ & $\mp0.01$  & $\mp0.05$ & $0.03$  & $0.04$ &  0.08 \\
                        \ion{K}{i}    & 769.89 & A & $\pm0.07$ & $\mp0.00$  & $\mp0.14$ & $0.04$  & $0.02$ &  0.16 \\
                        &                      & B & $\pm0.07$ & $\mp0.00$  & $\mp0.14$ & $0.04$  & $0.03$ &  0.16 \\
                        &                      & C & $\pm0.07$ & $\mp0.00$  & $\mp0.14$ & $0.04$  & $0.05$ &  0.17 \\
                        \hline
                \end{tabular}
        \end{center}
        The sign $\pm$ or $\mp$ reflects the change in the elemental abundance which occurs due to increase (top sign) or decrease (bottom sign) in a given atmospheric parameter. For example, increase in the effective temperature leads to an increase in the abundance ($\pm$) while increasing microturbulence velocity results in decreasing abundance ($\mp$).
        \vspace{-5mm}
\end{table*}

\subsection{Errors due to uncertainties in the atmospheric parameters}\label{sect_app:err_param}

The error in the determination of effective temperature, was estimated by computing effective temperatures using $V-I$ color indices that were increased/decreased by the amount given by their observational errors. For the latter, we used a conservative estimate of $\sigma(V)=\sigma(I)=0.03$\,mag. The obtained error in the effective temperature, $\pm65$\,K, was used to evaluate the influence of the uncertainty in \Teff\ on the determined abundances of Na, Mg, and K using the usual procedure adopted for this. The obtained errors, $\sigma(\Teff)$, are provided for all three elements in column~4, Table~\ref{tab_app:abund_err}.

The error in the effective temperature, along with those in luminosity ($\pm 0.03~{\rm L}_{\odot}$, estimated from photometric error in $M_{\rm V}$) and stellar mass ($\pm0.01~{\rm M}_{\odot}$, obtained from the isochrones), were used to obtain the error in $\log g$, $\pm0.04$\,dex. Although we used the slightly higher (and, in our view, more realistic) value of $\pm0.1$\,dex when estimating the resulting errors in the elemental abundances, $\sigma(\log g)$, we stress that these errors are very small and thus their contribution to the final abundance error is essentially negligible (Table~\ref{tab_app:abund_err}, column 5).

To estimate the error arising due to uncertainty in the microturbulence velocity, we utlized a sample of RGB stars in 47~Tuc for which $\xi_{\rm t}$ was determined spectroscopically in earlier studies using \ion{Fe}{i} lines. For this, we used 58 RGB stars from \citet{CBG09a} and 81 RGB stars from \citet{CPJ14} which were selected to match the same range in effective temperature as stars used in our sample. We then computed the mean microturbulence and its RMS variation, $\xi_{\rm t}=1.48\pm0.18$\,km/s which agrees well with $\xi_{\rm t}=1.5$\,km/s used in our study. The obtained RMS variation of microturbulence velocity was used as a representative uncertainty in $\xi_{\rm t}$ to estimate the resulting errors in the abundances of Na, Mg, and K, $\sigma(\xi_{\rm t})$ (Table~\ref{tab_app:abund_err}, column~6).

\subsection{Errors due to uncertainties in the continuum determination and line profile fitting}\label{sect_app:quality}

All spectral lines used in this study were carefully inspected for blends and/or possible contamination with telluric lines. Nevertheless, even the lines that were deemed to be the cleanest and thus suitable for the abundance analysis differed in their quality significantly. In order to take this into account, we grouped spectral lines of Na, Mg, and K into three classes according to their quality:

 \begin{list}{$\bullet$}{}
        \item A-class: strong or moderately strong lines with well-resolved line profiles;
        \item B-class: moderately blended lines, or lines that were insufficiently resolved in the line wings;
        \item C-class: weak, poorly resolved, or significantly blended lines.
 \end{list}

We then selected several representative spectra with lines of each class and identified wavelength intervals in the vicinity of Na, Mg, and K lines that were clean from spectral lines, either stellar or telluric (in fact, these intervals were used to determine continuum level for the abundance analysis). We further assumed that intensity variation in these intervals was caused entirely by the noise in the spectra. The error in the continuum determination was then computed as:

\begin{equation}
err(\rm cont)=\frac{\sigma_{\rm cont}}{\sqrt{N}} 
\label{scont}
,\end{equation}

\noindent where $\sigma_{\rm cont}$ is dispersion in the continuum variation obtained using all wavelength intervals available for the continuum determination in case of a given spectral line, and $N$ is the total number of wavelength points in these wavelength intervals. The resulting errors were used to estimate their impact on the abundance determination, that is, line profile fitting was done again with the continuum level shifted up and down by \textit{err(\rm cont)}, to obtain errors in the abundances of Na, Mg, and K. The resulting abundance errors are listed in column~7 of Table~\ref{tab_app:abund_err}.

To estimate the errors in the line profile fitting, we used the representative spectra selected in the previous step, as well as synthetic line profiles that were obtained during the abundance analysis as best-fits to these particular observed lines of Na, Mg, and K. We then computed RMS deviations between the observed and synthetic line profiles which were further converted into the uncertainties in the line equivalent width, and, finally, the resulting errors in the determined abundances, $\sigma_{\rm fit}$ (listed in column~8 of Table~\ref{tab_app:abund_err}).

\subsection{Total error in the abundances of Na, Mg, and K}

The individual contributions to the total error due to uncertainties in the atmospheric parameters, continuum determination, and spectral line profile fitting were added in quadratures to obtain the total error in the abundances of Na, Mg, and K. These errors are provided in the last column of Table~\ref{tab_app:abund_err}. We stress that they provide a lower limit for the uncertainties in the determined abundances of Na, Mg, and K since they do not account for various systematic uncertainties that are unavoidable in the abundance analysis procedure.

\section{Maximum-likelihood testing of the intrinsic spread in elemental abundances}\label{sect_app:ML_test}

In order to estimate the size of the possible intrinsic spread in the abundances of Na, Mg, and K, we followed the procedure described in \citet{MBI12,MBM15} which was used by these authors to study K abundance spreads in NGC~2419 and NGC~2808. For this, maximum-likelihood (ML) technique was utilized to evaluate the mean abundance ratio, $\xtmean{\mbox{[A/B]}}$, of elements A and B, as well as its intrinsic spread, $\sigma_{\rm int}$\footnote{In this study [Na/Fe], [Mg/Fe], and [K/Fe] abundance ratios were used.} We defined the likelihood as

\begin{equation}
\label{eq:ML}
L(\xtmean{\mbox{[A/B]}}, \sigma_{\rm int}) = \prod_{\rm i=1}^{N}
p_{\rm i}(\xtmean{\mbox{[A/B]}}, \sigma_{\rm int})
,\end{equation}

\noindent where the multiplication was done using quantities computed for each star in a sample of $N$ stars. In the equation above, $p_{\rm i}$ is a Gaussian probability function defined as

\begin{table}
        \begin{center}
                \caption{Results of the maximum-likelihood (ML) testing of the intrinsic spread in the abundances of Na, Mg, and K.}\label{tab_app:ML_values}
                \begin{tabular}{lccc}
                        \hline\hline
                        Element & $\langle [X_{\rm i}/\ion{Fe}]\rangle$, likelihood value & $\sigma[X_{\rm i}/\ion{Fe}]$ & $\sigma_{\rm int}[X_{\rm i}/\ion{Fe}]$   \\ [0.5ex]
                        $X_{\rm i}$     &    dex                  &   dex    & dex             \\
                        \hline
                        Na        & $0.49\pm0.02$  & $0.13$ & $0.04\pm0.05$  \\
                        Mg        & $0.48\pm0.02$  & $0.11$ & $0.08\pm0.02$  \\
                        K         & $0.10\pm0.03$  & $0.14$ & $0.00\pm0.05$  \\
                        \hline
                \end{tabular}
                \label{err}
        \end{center}
        \vspace{-5mm}
\end{table}

\begin{equation}
\resizebox{0.45 \textwidth}{!}
{
$p_{\rm i}(\xtmean{\mbox{[A/B]}}, \sigma_{\rm int}) =
\frac{1}{\sqrt{\sigma_{\rm int}^2 + \sigma_{\rm i}^2}} {\rm exp}
\left [-\frac{1}{2} \left (\frac{\xtmean{\mbox{[A/B]}} -
        \xtmean{\mbox{[A/B]}}_{\rm i}}{\sqrt{\sigma_{\rm int}^2 + \sigma_{\rm
                        i}^2}} \right )^2\right ]$
}
\label{eq:pgauss}
\end{equation}

\noindent where $\xtmean{\mbox{[A/B]}}_{\rm i}$ and $\sigma_{\rm i}$ is the abundance ratio and its error for the \textit{i}-th star in the sample taken from Table~\ref{tab_app:all_abund}.

We computed likelihood values for a 2D grid that was defined by various combinations of parameters ($\xtmean{\mbox{[A/B]}}$, $\sigma_{\rm int}$). The mean abundance and its internal spread was determined by maximizing likelihood values in the computed 2D grid. We also computed errors in the derived quantities according to the prescription given in \citet[][see their equations (5)--(10)]{PM93}.

The obtained mean abundance ratio, $\langle [X_{\rm i}/\ion{Fe}]\rangle$, and its~uncertainty, as well as the total variation, \mbox{$\sigma[X_{\rm i}/\ion{Fe}]$~($\equiv\sqrt{\sigma_{\rm int}^2 + \sigma_{\rm i}^2}$)}, and the determined intrinsic variation, $\sigma_{\rm int}[X_{\rm i}/\ion{Fe}]$, in the abundance of element $X_{\rm i}$ are provided in Table~\ref{tab_app:ML_values}. Although our test results suggest the existence of small intrinsic spread of similar size in case of Na and Mg, we note that for Mg the real spread may in fact be significantly smaller (we note that although we used abundance-to-iron ratios, iron abundance was assumed to be fixed and thus it had no effect on the spread of individual abundance-to-iron ratios). The reason for this is that the spectral lines of \ion{Mg}{i} used in our work were relatively poorly resolved in the \twodFH\ spectra which in many cases meant that only part of the observed line profile could be fitted with the synthetic profile. This  made the total error of the line profile fitting smaller which, in turn, may have lead to an  overestimated intrinsic abundance variation. Therefore, existence of the intrinsic spread in the case of Mg needs to be verified using higher quality spectroscopic data.

\end{appendix}

\end{document}